\documentclass[twocolumn,showpacs,showkeys,amsmath,amssymb,page]{revtex4}
\usepackage{graphicx}% Include figure files
\usepackage{dcolumn}% Align table columns on decimal point
\usepackage{bm}% bold math

\begin{document}

%\preprint{Draft version 4r 19:20}
\preprint{}

\title{Evolutionary Stability of Ecological Hierarchy}
% Force line breaks with \\

\author{Taksu Cheon}
\email[E-mail:]{taksu.cheon@kochi-tech.ac.jp }
\homepage[\\]{http://www.mech.kochi-tech.ac.jp/cheon/}
\affiliation{
Laboratory of Physics,
Kochi University of Technology,
Tosa Yamada, Kochi 782-8502, Japan
}

%\date{\today}
\date{January 31, 2003}

\begin{abstract}
A self-similar hierarchical solution that is both
dynamically and evolutionarily stable is found to the 
multi dimensional Lotka-Volterra equation with a single chain of
prey-predator relations. This gives a simple and natural
explanation to the key features of hierarchical
ecosystems, such as its ubiquity, pyramidal population
distribution, and higher aggressiveness among higher trophic levels.
\end{abstract}

\pacs{87.23.Kg, 89.75.Da, 05.45.-a}
\keywords{Lotka-Volterra equation, 
Trophic pyramid, Self-similarity}
%Use showkeys class option if keyword
\maketitle

%
%\section{Introduction}
%

From bacterial colony to human society, hierarchical structure
is one of the most universal features of ecological systems.
It is typically realized as a {\it trophic pyramid} with exponentially 
larger populations for lower trophic levels. 
The ubiquity of this hierarchy in nature suggests the existence
of a simple and robust mechanism behind it.

Let us consider an ecosystem of $N$ species whose
populations $x_1(t)$, $x_2(t)$, $...$, $x_N(t)$ are described by a
set of coupled differential equations with parameters
that represent the environmental conditions \cite{MA72,HS88}.
Then, the robustness of ecological structures
is expressed as the presence and stability
of fixed point solutions.
The stability should be considered on two levels of different time-scales.  
The short-term dynamical stability is the stability against the 
perturbation in the dynamical variables $x_n$, 
while the long-term stability is related
to the robustness of the solutions against the secular variation of
environmental parameters.
When some environmental parameters are at the disposal
of a certain species, 
natural selection will lead to the realization of the parameter 
value that maximizes the population of that species.  
This is the concept of evolutionary stability \cite{MS82,AX97}.

There are several numerical simulations of multi-species 
ecosystems that show the spontaneous emergence of
multi-trophic structure  \cite{LA01,QH02}.   
The models employ coupled
differential equations with stochastic parameter
variation subjected to evolutionary selection rules. 
These results indicate that pyramidal hierarchy 
%with exponentially abundant lower trophic populations 
is an evolutionarily stable configuration of ecosystems
irrespective to the fine detail of the model.
It is high time to search for a simple and clear theoretical 
explanation before further numerical calculations 
with increasingly ``realistic'' settings are to be pursued.

In this article, we consider ecosystems modeled by the
Lotka-Volterra equation describing $N$ species that form a single
vertical chain of prey-predator relations.  We intend to prove the
existence of hierarchical solutions that are stable both 
dynamically and evolutionarily.

%
%\section{Aggressiveness as Evolutionarily Optimized Parameter}
%

Let us begin with the two species 
prey-predator Lotka-Volterra equation
%
% 1
\begin{eqnarray}
\label{lv2}
 \dot{x_1} &=& b x_1 - a x_1^2 - \rho_2 x_1 x_2
\\ \nonumber
\dot{x_2} &=& -d_2 x_2 + f_2 \rho_2 x_1 x_2 .
\end{eqnarray}
Here $b$ is the reproduction rate of the prey $x_1$ 
and $a$ the environmental limitation factor to its growth. 
The parameter $d_2$ is the decay rate of 
the predator $x_2$ which will not subsist without preying 
on $x_1$ with the rate of aggression $\rho_2$. 
The factor $f_2$ represents the combination of 
the efficiency of the predation and the average 
mass ratio between prey and predator individuals.
All parameters are positive real numbers.
%We treat all parameters apart from $\rho_2$ 
%as the environmentally given fixed parameter.
Although the dynamical variables $x_n$ are 
treated as continuous quantities here, they are
approximations of the actual integer populations.
Moreover, in real life, there is a threshold number
for a population under which a species is not viable.
As is immediately identified, (\ref{lv2}) 
has a nontrivial fixed point solution $x_1(t) = X_1$, 
$x_2(t) = X_2$ where $X_1$ and $X_2$ satisfy
%
% 2
\begin{eqnarray}
\label{lv2fix} b - a X_1 - \rho_2 X_2 &=& 0 ,
\\ \nonumber
-d_2 + f_2 \rho_2 X_1  &=& 0 .
\end{eqnarray}
The eigenvalues $\lambda$ of the linearized map 
around the fixed point are given by
%
% 3
\begin{eqnarray}
\label{lv2lin}
%\lambda = -{ {ad_2} \over {2f_2 \rho_2} }
% \left
%  ( 1 \pm 
%    \sqrt{ 1+ { {4f_2\rho_2}\over{a} }(1-{{f_2 b\rho_2}\over {ad_2}})}
%  \right) .
%\end{eqnarray}
\lambda = -{ {1} \over {2 q} }
  \left( 1 \pm \sqrt{ 1- 4d_2q(bq-1)}
  \right) ,
\quad
q\equiv{f_2\rho_2 \over {ad_2} }.
\end{eqnarray}
Therefore, the fixed point is dynamically stable when 
we have $f_2\rho_2b$ $>ad_2$.  Since, in this work, we are primarily 
concerned with the ``populous'' regime, $b \gg a$, 
this condition is almost always satisfied.

Let us now assume that the aggression parameter $\rho_2$ is a
quantity that is at the disposal of the predator $x_2$ through
a long term ``behavioral change''.
That is, we regard $X_1$ and $X_2$ as functions of $\rho_2$. 
Naturally, a change in $\rho_2$ would be
directed toward the maximization of the predator population $x_2$
through evolutionary selection.
We further assume that the time scale for the change of $\rho_2$ is
substantially larger than the time scale for the variation of $x_1(t)$ 
and $x_2(t)$. Then, irrespective to the
precise mechanism of the variation of $\rho_2$, one eventually ends
up with the value $\rho_2^\star$ that maximizes $X_2(\rho_2)$. 
With the notation $X_i^\star$ $\equiv X_i(\rho_2^\star)$, we have
%
% 4
\begin{eqnarray}
\label{optima2}
\rho_2^\star = { {2 a d_2} \over { f_2 b } },
\quad X_2^\star
={{f_2b^2} \over {4 a d_2} } ,
\quad X_1^\star = { b \over {2a}}  ,
\end{eqnarray}
which represents the evolutionarily stable solution.
The solution is always dynamically stable, 
since, at these values, the real part of the eigenvalue of linearized 
map (\ref{lv2lin})  never becomes positive. 
The stability against parametric variation of $\rho_2$ can
be judged by
%
% 5
\begin{eqnarray}
\label{curv2}
 { {d^2 X_2^\star} \over {d \rho_2^2}} 
 = -{ b \over { {\rho_2^\star}^3 } }.
\end{eqnarray}
Our result shows that the optimal aggression rate from the 
stand point of the predator is to hunt the prey down to one half 
of its natural stability point $b/a$ that is reached by $x_1$ 
when left alone. Already at this point, (\ref{optima2}) 
gives us some insights. 
When $b$ and $d_2$ are comparable quantities,
the predator population $X_2^\star$ is 
suppressed by the factor $f_2/2$
compared to the prey $X_1^\star$.
Since $f_2$ is typically smaller than 1, we tend to have
a small number of predators supported by a large pool
of prey biomass as a stable configuration. 
Another interesting point is that 
the milder environment signified by a higher value of
$b/a$ will increase both $X_1^\star$ 
and $X_2^\star$ while reducing the optimal aggression
rate $\rho_2^\star$ of the predator.
This principle of {\it noblesse oblige} is a widely observed,
but nonetheless nontrivial aspect of life. 
This is corroborated, for example, by recent field work
observation on slave-making ants \cite {FH01}.

According to (\ref{optima2}), the only way 
for the prey $x_1$ to increase 
its equilibrium population is to ``improve the environment'' by 
increasing $b/a$, when there is any such mean available to it. 
An intriguing fact is that ``improving defensive shield'' by 
decreasing $f_2$ will not benefit $x_1$ directly;
it simply decreases the predator population $X_2^\star$.  
However, when $X_2$ is close to the viability threshold, 
a decreasing $f_2$ would be a sensible strategy for $x_1$, 
since that could drive $x_2$ out of existence, which would 
result in the instant doubling of $X_1^\star$.
%%%

%
%\section{Tritrophic Prey-Predator Model}
%

Next, we consider the case of $N=3$ species that forms a single
chain of prey-predator relations \cite{CP02}:
%
% 6
\begin{eqnarray}
\label{lv3} 
\dot{x_1} &=& b x_1 - a x_1^2 - \rho_2 x_1 x_2 ,
\\ \nonumber
\dot{x_2} &=& -d_2 x_2 + f_2 \rho_2 x_1 x_2 -\rho_3 x_2 x_3 ,
\\ \nonumber
\dot{x_3} &=& -d_3 x_3+ f_3 \rho_3 x_2 x_3 .
\end{eqnarray}
The fixed point solution is obtained as
%
% 7
\begin{eqnarray}
\label{lv3sol}
b  - a X_1 - \rho_2 X_2 &=& 0 ,
\\ \nonumber
-d_2 + f_2 \rho_2 X_1  -\rho_3  X_3 &=& 0 ,
\\ \nonumber
-d_3 + f_3 \rho_3 X_2  &=& 0.
\end{eqnarray}
By rearranging the first two equations, 
we obtain
%
% 8
\begin{eqnarray}
\label{lv3eff2}
b_2  - a_2 X_2 - \rho_3 X_3 &=& 0 ,
\\ \nonumber
-d_3 + f_3 \rho_3 X_2  &=& 0.
\end{eqnarray}
with
%
% 9
\begin{eqnarray}
\label{a2b2}
a_2 \equiv {{f_2 \rho_2^2} \over a},
\quad
b_2 \equiv {{f_2 b \rho_2} \over {a}} - d_2.
\end{eqnarray}
The problem is therefore reduced to the $N = 2$ case with the
predator $x_3$ and effectively self-sustaining prey $x_2$ which
has reproduction and limiting coefficients $b_2$ and $a_2$. If the
top predator, driven by evolutionary selection, tries to 
maximize its equilibrium population $X_3$ by varying $\rho_3$,
it will reach the optimum given by
%
% 10
\begin{eqnarray}
\label{optima3}
\rho_3^{(\star)}= { { 2 a_2 d_3} \over {f_3 b_2}  },
\quad
X_3^{(\star)}
={{f_3 b_2^2} \over {4 a_2  d_3} } ,
\quad
X_2^{(\star)} =
{ {b_2} \over {2 a_2 }} .
\end{eqnarray}
The fact that these values are optimum only with a given $\rho_2$
is indicated by the bracketed asterisk.  
From the relation $\rho_3^{(\star)}X_3^{(\star)}$
$=b_2/2$, one can rewrite the first two equations of (\ref{lv3sol}) as
%
% 11
\begin{eqnarray}
\label{rec1}
{b \over 2}  - a \left( { X_1 - {b\over{2a}} } \right) - \rho_2 X_2 &=& 0
\\ \nonumber
-{d_2 \over 2} + f_2 \rho_2 \left( { X_1 - {b\over{2a}} } \right)   &=& 0.
\end{eqnarray}
This is essentially the same relationship as in the $N=2$ case
(\ref{lv2fix}), with an extra factor $1/2$ in front of the first terms,
and the shift in $X_1$ in the socond.  We should now suppose that 
the middle predator $x_2$ will, in a long run, adjust its aggression
rate $\rho_2$ toward the prey $x_1$ and maximize $X_2$.
We then obtain the solution
%
% 12
\begin{eqnarray}
\label{optima32}
\rho_2^\star = { {2 a d_2} \over { f_2 b } } ,
\quad
X_2^\star ={ {f_2b^2} \over {8 a d_2} } ,
\quad
X_1^\star ={ {3b} \over {4 a} } ,
\end{eqnarray}
which in turn yields
%
% 13
\begin{eqnarray}
\label{optima33}
\rho_3^\star
= { { 8 a d_2 d_3} \over  {f_2 f_3 b^2} },
\quad
X_3^\star = { f_2 f_3 b^2 \over {16 a d_3}}.
\end{eqnarray}
Note the fact that $\rho_2^\star$ here is
identical to the $N=2$ case.
We also obtain parametric stability measures as
%
% 14
\begin{eqnarray}
\label{curv3}
 { {d^2 X_2^\star} \over {d \rho_2^2}} 
 = -{ b \over { 2{\rho_2^\star}^3 } },
 \quad
  { {d^2 X_3^\star} \over {d \rho_2^2}} 
  = -{ {d_2} \over { {\rho_3^\star}^3 } } ,
\end{eqnarray}
which indeed prove the evolutionary stability of the solution.
%
%%%%%%%%%% fig.1
% 
\begin{figure}
\includegraphics{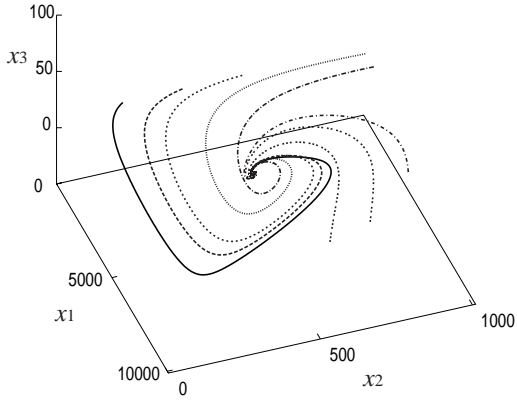}% Here is how to import EPS art
\caption{\label{fig:epsart}
The phase space profile of a 
$N=3$ system at evolutionarily stable parameter value. 
The parameters are set to be $b=1$, $a=1/10000$, $d_2=1/3$,
$d_3=1/5$, $f_2=1/10$, $f_3=1/5$.
The optimal aggression rates are calculated as
 $\rho_2^\star=1/1500$ and $\rho_3^\star=1/375$.
All orbits approach 
the fixed point $(X_1^\star, X_2^\star, X_3^\star)$ $ =(7500,375, 64.5)$. 
}
\end{figure}
%
%%%%%%%%%% fig.1
%
In Fig.1, the phase space profile of one such example 
of evolutionarily stable $N=3$ system is depicted.
With consideration at each stage (\ref{lv3eff2}) and (\ref{rec1}),
it is easy to see that this evolutionarily stable solution is also
dynamically stable for all parameter values. In
effect, the single-chain $N = 3$ Lotka-Volterra equation is
broken into two $N = 2$ equations with essentially the same
structure, albeit with an additional factor for the lower chain.

When $d_2$ and $d_3$ are comparable quantities,
the population of the top trophic level $X_3^\star$ is 
inherently suppressed by the factor $f_3/2$
compared to that of $X_2^\star$, giving a pyramidal
profile to the trophic structure.
It is amusing to note that, from the stand point of
the lowest trophic species, an $N=3$ system, in which
two thirds of its natural population is left alive,
is considerably more ``benign'' than an $N=2$ system. 

%%%
%\section{Recursive Solution of Hierarchical Prey-Predator Model}
%

The preceding proof for the $N=3$ solution suggests 
its generalization to arbitrary $N$.  
This is achieved through the realization 
that the fixed point equation
for any mid-level population $x_n$ can have both purely prey-like
and purely predator-like representations.  Let us start with the
$N$ vertically-coupled Lotka-Volterra equation with evolutionarily
adjustable aggression parameter $\rho_n$ for each species $x_n$
%
% 15
\begin{eqnarray}
\label{lvN}
\dot{x}_1 &=& b x_1 - a x_1^2 - \rho_2 x_1 x_2 ,
\\ \nonumber
\dot{x}_2 &=&-d_2 x_2 + f_2 \rho_2 x_{1} x_2 - \rho_{3} x_2 x_{3} ,
\\ \nonumber
\vdots
\\ \nonumber
\dot{x}_n &=&-d_n x_n + f_n \rho_n x_{n-1} x_n - \rho_{n+1} x_n x_{n+1} ,
%\\ \nonumber 
%& & \qquad\qquad\qquad\qquad\qquad (n=2, ..., N-1)
\\ \nonumber
\vdots
\\ \nonumber
%\dot{x}_{N-1} &=&-d_{N-1} x_{N-1} 
%+ f_{N-1} \rho_{N-1} x_{N-2} x_{N-1} 
%- \rho_{N} x_{N-1}x_{N} ,
%\\ \nonumber 
\dot{x}_N &=& -d_N x_N + f_N \rho_N x_{N-1} x_N  .
\end{eqnarray}
In general, a trophic level can comprise several competing species.
In our simplified treatment, however, such species are lumped into 
a single population variable.  
The equations for the nontrivial fixed point $(X_1, ..., X_N)$ are
%
% 16
\begin{eqnarray}
\label{lvNf}
 b  - a X_1 - \rho_2  X_2 &=& 0,
\\ \nonumber
-d_2 + f_2 \rho_2 X_{1} - \rho_{3} X_{3} &=& 0,
\\ \nonumber
&\vdots&
%&:&
\\ \nonumber
-d_n + f_n \rho_n X_{n-1} - \rho_{n+1} X_{n+1}&=& 0, 
\\ \nonumber
&\vdots&
%&:&
\\ \nonumber
-d_N + f_N \rho_N X_{N-1} &=& 0 .
\end{eqnarray}
Apart from the species with the highest trophic level  $x_N$, each of
these can be transformed to the form
%
% 17
\begin{eqnarray}
\label{slavN}
 b_{n} - a_{n} X_n  - \rho_{n+1} X_{n+1} &=& 0 ,
\end{eqnarray}
with the recursive definition 
%
% 18
\begin{eqnarray}
\label{aNbN} 
a_{n} = f_{n} \rho_{n}^2 { 1 \over {a_{n-1}} }  ,
\quad
b_{n} = f_{n} \rho_{n} {{b_{n-1}}\over{a_{n-1}}} - d_n .
\end{eqnarray}
%
%It can be further rewritten into 
Rewriting (\ref{slavN}), we obtain the ``slave''  form
%
% 19
\begin{eqnarray}
\label{slavN2}
  \eta_n b_{n} - a_{n} \left( X_n - (1-\eta_n) {b_n \over a_n} \right)
 - \rho_{n+1} X_{n+1} = 0
%\\ \nonumber
% (n=1, .., N-1).
\end{eqnarray}
for $n=1, .., N-1$.
Let us now {\it assume} the relation
%
% 20
\begin{eqnarray}
\label{mastern}
 \rho_{n+1}^{(\star)} X_{n+1}^{(\star)} &=& {{\eta_n b_n}\over 2} .
%\\ \nonumber
%&=& -{{\eta_n}\over 2}d_n + {{\eta_n f_n \rho_n}\over {2}}
% {b_{n-1} \over a_{n-1}}
\end{eqnarray}
%
%% v5
%This assumption results in another form 
%of (\ref{lvNf}):
%for the fixed point equations
%(\ref{lvNf}) apart from the lowest trophic level:
%
% old eq twenty one
%\begin{eqnarray}
%\label{masterN}
%%-\left( 1-{\eta_{n+1} \over 2}\right)d_{n+1}
%-{{ 2-\eta_{n+1} } \over 2} d_{n+1}
%%\\ \nonumber
%+ f_{n+1} \rho_{n+1}
%\left( X_{n} - {\eta_{n+1}\over 2} {b_{n} \over a_{n}} 
%\right)  = 0 .
%%\\ \nonumber
%%(n=1, .., N-1),
%\end{eqnarray}
%
If we combine (\ref{lvNf}), (\ref{mastern}) with a requirement
%
% 21
\begin{eqnarray}
\label{hrec} 
1-\eta_{n} = {{\eta_{n+1}}\over 2} ,
\end{eqnarray}
we obtain the ``master'' form
%
% 22
\begin{eqnarray}
\label{masterN2}
-\eta_{n}d_{n+1} + f_{n+1} \rho_{n+1}
\left( X_n - (1-\eta_n) {b_n \over a_n} \right) = 0
%\\ \nonumber
% (n=1, .., N-1).
\end{eqnarray}
for $n=1, .., N-1$. 
The equations (\ref{lvNf}) are now decoupled to $(N-1)$ pairs of 
prey-predator equations (\ref{slavN2}) and (\ref{masterN2}).  
We then have
%
% 23
\begin{eqnarray}
\label{solvN}
 \rho_{n+1}^{(\star)} = {2a_{n} d_{n+1} \over {f_{n+1} b_{n}}} ,
\end{eqnarray}
and
%
% 24
\begin{eqnarray}
\label{solvN2}
 X_{n+1}^{(\star)} = \eta_{n} { {f_{n+1} b_{n}^2} \over {4 a_{n} d_{n+1}}} ,
\quad
 X_{n}^{(\star)} = (2-\eta_{n}) { {b_{n}} \over {2 a_{n} } } .
\end{eqnarray}
This result justifies the assumption (\ref{mastern}) {\it a posteriori}, 
and the whole procedure becomes consistent.
From the last equation of (\ref{lvNf}), 
we observe that $\eta_{N-1}$ should be set to one,
which results in $\eta_{N-2} = 1/2$,
$\eta_{N-3} = 3/4$, $\cdots$.
We finally obtain the following explicit forms
for the evolutionarily and dynamically stable solution:
%
% 25
\begin{eqnarray}
\label{optimaN}
X_1^\star &=& { B_{N} b \over {2^{N-1} a}},
\\ \nonumber
X_2^\star &=& { B_{N-1} f_2 b^2 \over {2^{N} a d_2}},
\quad
\rho_2^\star = {  { 2 a d_2} \over {f_2 b} },
\\ \nonumber
%X_3^\star &=& { B_{N-2} f_2 f_3 b^2 \over {2^{N+1} a d_3}},
%\quad
%\rho_3^\star = { { 8 a d_2 d_3} \over {f_2 f_3 b^2} },
%\\ \nonumber
&\vdots&
\\ \nonumber
X_N^\star &=&
{ B_{1} f_2 \cdots f_N b^2 \over {2^{2N-2} a d_3}},
 \quad
\rho_N^\star =
{ { 2^{2N-3} a d_2 d_3} \over {f_2 \cdots f_N b^2} } .
\end{eqnarray}
The stability with respect to the variation $\rho_n$ is given by
%
% 26
\begin{eqnarray}
\label{optimaN2}
{ {d^2 X_2^\star} \over {d \rho_2^2}} =
-{ B_{N-1}b \over { 2^{N-2}{\rho_2^\star}^3 } },
\quad
%\\ \nonumber
{ {d^2 X_3^\star} \over {d \rho_3^2}} =
-{ B_{N-2}b \over { 2^{N-3}{\rho_3^\star}^3 } },
\\ \nonumber
\cdots
\quad
%\\ \nonumber
{ {d^2 X_N^\star} \over {d \rho_N^2}} =
 -{ B_{1}b \over { {\rho_N^\star}^3 } }.
\end{eqnarray}
Here the coefficient $B_{n}$ is a variant of the Fibonacci series
defined by
%
%27
\begin{eqnarray}
B_{n+2} = B_{n+1} + 2 B_{n}, \quad B_1 = B_2 = 1.
\end{eqnarray}
Some of the numbers are $B_3 = 3$, $B_4 = 5$, $B_5 = 11$, $B_6 =
21$, $B_7=43$, $\cdots$.
%

%
%%%%%%%%Table I
%
\begin{table*}
\label{tab1 }
\caption{
The evolutionarily stable hierarchical population for  $N$ species Lotka-Volterra
equation up to $N = 5$. 
$X_n^\star$ is the population of $n$-th trophic level, and $\rho_n^\star$ 
its aggression rate toward its prey. 
               }
\begin{tabular}{ccc}
% after \\ : \hline or \cline{col1-col2} \cline{col3-col4} ...
   & $N=1$ &  \\
 & & \\
\hline
 $n$   & $X_n^\star$ & $\rho_n^\star$  \\
\hline
  & & \\
1 & {\Large ${b\over{a}}$ } &  \\
  & & \\
\hline
\end{tabular}
\quad
\begin{tabular}{ccc}
% after \\ : \hline or \cline{col1-col2} \cline{col3-col4} ...
   & $N=2$ &  \\
 & & \\
\hline
 $n$   & $X_n^\star$ & $\rho_n^\star$  \\
\hline
  & & \\
2 & {\Large ${  f_2 b^2 \over {4 a d_2}}$ } &
 {\Large ${ { 2 a d_2} \over {f_2 b} }$ } \\
  & & \\
1 & {\Large ${b\over{2a}}$ } & \\
  & & \\
\hline
\end{tabular}
\quad
\begin{tabular}{ccc}
% after \\ : \hline or \cline{col1-col2} \cline{col3-col4} ...
   & $N=3$ &  \\
  & & \\
\hline
 $n$   & $X_n^\star$ & $\rho_n^\star$  \\
\hline
  & & \\
 3 & {\Large ${  f_2 f_3 b^2 \over {16 a d_3}}$ } &
  {\Large ${ { 8 a d_2 d_3} \over {f_2 f_3 b^2} }$ }\\
  & & \\
 2 & {\Large ${  f_2 b^2 \over {8 a d_2}}$ } &
  {\Large ${ { 2 a d_2} \over {f_2 b} }$ } \\
  & & \\
 1 & {\Large ${3b \over{4a}}$ } & \\
  & & \\
\hline
\end{tabular}
\quad
\begin{tabular}{ccc}
% after \\ : \hline or \cline{col1-col2} \cline{col3-col4} ...
   & $N=4$ &  \\
  & & \\
\hline
 $n$   & $X_n^\star$ & $\rho_n^\star$  \\
\hline
  & & \\
 4 & {\Large ${  f_2 f_3 f_4 b^2 \over {64 a d_4}}$ }&
{\Large ${ { 32 a d_3 d_4} \over {f_2 f_3 f_4 b^2} }$ }\\
  & & \\
 3 & {\Large ${  f_2 f_3 b^2 \over {32 a d_3}}$ } &
{\Large ${ { 8 a d_2 d_3} \over {f_2 f_3 b^2} }$ }\\
  & & \\
 2 & {\Large ${ 3 f_2 b^2 \over {16 a d_2}}$ } &
{\Large ${ { 2 a d_2} \over {f_2 b} }$ }\\
  & & \\
 1 & {\Large ${5b \over{8a}}$ }& \\
  & & \\
\hline
\end{tabular}
\quad
\begin{tabular}{ccc}
% after \\ : \hline or \cline{col1-col2} \cline{col3-col4} ...
   & $N=5$ &  \\
  & & \\
\hline
 $n$   & $X_n^\star$ & $\rho_n^\star$  \\
\hline
  & & \\
 5 &{\Large $ {  f_2 f_3 f_4 f_5 b^2 \over {256 a d_5} }  $ }&
{\Large ${ { 128 a d_4 d_5} \over {f_2 f_3 f_4 f_5 b^2} }$ }\\
  & & \\
 4 &{\Large ${  f_2 f_3 f_4 b^2 \over {128 a d_4}}$ }&
{\Large ${ { 32 a d_3 d_4} \over {f_2 f_3 f_4b^2} }$ }\\
  & & \\
 3 & {\Large ${  3 f_2 f_3 b^2 \over {64 a d_3}}$ }&
{\Large ${ { 8 a d_2 d_3} \over {f_2 f_3 b^2} }$ }\\
  & & \\
 2 &{\Large ${  5 f_2 b^2 \over {32 a d_2}}$ }&
{\Large ${ { 2 a d_2} \over {f_2 b} }$ }\\
  & & \\
 1 & {\Large ${11b\over{16a}}$ }& \\
  & & \\
\hline
\end{tabular}
\end{table*}
%
%%%%%%%%Table
%
%
In table I, the hierarchical solutions up to $N=5$ are listed.   
The most notable feature is of course the exponentially smaller
population in higher trophic levels.  
Assuming $d_n \approx b$ for
all $n$, we have a decrease in the population by factor $f_n/2$ for
each increase of one trophic level.  Since $f_n$ is in general
substantially smaller than one, we get a pyramidal hierarchy with
a steep exponential decrease. 
We should also mention the self-similarity
of the solution:
For any given trophic level, the portion of its ``natural'' population
saved  from exploitation by higher trophic levels 
varies like $1/2$, $3/4$, $5/8$, $\cdots$, 
whenever more trophic levels are added on top.
On the other hand,  its optimal aggression rate is 
unaffected by the presence of higher trophic levels.
A higher value of $\rho_n^\star$ for larger $n$ is a direct result of
the scarcity of its prey.

Ultimately, the quantity $b/a$ gives the base biomass, on top of 
which the whole trophic pyramid structure is built.  
Since there is a minimum population
for the highest trophic species to be viable,
this naturally puts a limit to the maximum number 
for the trophic hierarchy of an ecosystem with a given base biomass.

%
%\section{Summary and Prospects}
%

In summary, within the framework of 
a single vertical food chain model,
%% v5
a pyramidal self-similar hierarchy is found
in Lotka-Volterra system.
It might be possible to generalize our results to models
with plural species in each trophic levels.   Hopefully,
the search for generic properties of Lotka-Volterra
system along this line shall provide a solid 
backbone for experimental and numerical studies of ecosystems.

\medskip
%
%\begin{figure*}
%\includegraphics{fig_2.eps}% Here is how to import EPS art
%\includegraphics{fig_2.pdf}% Here is how to import EPS art
%\caption{\label{fig:wide}Use the figure* environment to get a wide
%figure that spans the page in \texttt{twocolumn} formatting.}
%\end{figure*}
%
%\begin{acknowledgments}
%
The author wishes to
%% v5 acknowledge his gratitude to 
thank Izumi Tsutsui,
Takuma Yamada, Koji Sekiguchi and David Greene 
for helpful discussions and useful comments.
%\end{acknowledgments}

%\appendix

%\section{Appendixes}


\begin{thebibliography}{99}
%------------------%
\bibitem{MA72}
R.M. May,
%%v5 {\em Simple mathematical models with very complicated dynamics},
Nature,
{\bf 261}, 459-467 (1976).
%
%\bibitem{MA74}
%R.M. May,
%{\em Stability and complexity in model ecosystems},
%Priceton Univ. Press, Princeton, 1974).
%
\bibitem{HS88}
J. Hofbauer and K. Sigmund: 
{\em The Theory of Evolution and Dynamical Systems}, 
(Cambridge Univ. Press, 1988).
%
\bibitem{MS82}
J. Maynard Smith,
{\em Evolution and the theory of games},
(Cambridge Univ. Press, Cambridge, 1982).
%
\bibitem{AX97}
R. Axelrod,
{\em The Complexity of Cooperation:
Agent-Based Models of Competition and Cooperation}
(Princeton Univ. Press, 1997). 
%
\bibitem{LA01}
M. Lassig, U. Bastolla, S.C. Manrubia and A. Valleriani,
%%v5 {\em Shape of ecological networks},
Phys. Rev. Lett.
{\bf 86}, 4418-4421 (2001).
%
\bibitem{QH02}
C. Quince, P.G. Higgs and A.J. McKane,
%%v5 {\em Food web structure and  the evolution of ecological communities},
arXiv.org, nlin.AO/0105057 (2001).
%
%\bibitem{MA00}
%J.S.S. Martins,
%{\em Simulated coevolution in a mutating ecology},
%Phys. Rev.  {\bf E61} (2000) 2212.
%
\bibitem{FH01}
S. Foitzik, C.J. DeHeer, D.N. Hunjan and J.M. Herbers,
%%v5 {\em Coevolution in host-parasite systems:
%% Behavioural strategies of slave-making ants and their hosts},
Proc.  Roy. Soc. London {\bf B268}, 1139-1146 (2001) .
%
\bibitem{CP02}
E. Chauvet, J. Paullet, J.P. Previte and Z. Walls,
%%v5 {\em A Lotka-Volterra three-species food chain},
Math. Magazine  {\bf 75},  243-255 (2002).
%
%------------------%
\end{thebibliography}
%------------------%

%------------------%
\end{document}